\documentclass[12pt]{article}
\usepackage{epsfig}
\usepackage{ifthen} 
\newboolean{uprightparticles}
\setboolean{uprightparticles}{false} 

\usepackage{lineno}  
\usepackage{xspace} 

\usepackage{amsmath} 
\usepackage{amssymb}
\usepackage{amsfonts}
\usepackage{upgreek} 

\newcommand*\patchAmsMathEnvironmentForLineno[1]{%
\expandafter\let\csname old#1\expandafter\endcsname\csname #1\endcsname
\expandafter\let\csname oldend#1\expandafter\endcsname\csname
end#1\endcsname
 \renewenvironment{#1}%
   {\linenomath\csname old#1\endcsname}%
   {\csname oldend#1\endcsname\endlinenomath}%
}
\newcommand*\patchBothAmsMathEnvironmentsForLineno[1]{%
  \patchAmsMathEnvironmentForLineno{#1}%
  \patchAmsMathEnvironmentForLineno{#1*}%
}
\AtBeginDocument{%
\patchBothAmsMathEnvironmentsForLineno{equation}%
\patchBothAmsMathEnvironmentsForLineno{align}%
\patchBothAmsMathEnvironmentsForLineno{flalign}%
\patchBothAmsMathEnvironmentsForLineno{alignat}%
\patchBothAmsMathEnvironmentsForLineno{gather}%
\patchBothAmsMathEnvironmentsForLineno{multline}%
}

\usepackage{hyperref}    
\usepackage[all]{hypcap} 

\newcommand{\Cpar}{\ensuremath{C}\xspace}
\newcommand{\Cbpar}{\ensuremath{C_{\bar{f}}}\xspace}
\newcommand{\Dpar}{\ensuremath{D_f}\xspace}
\newcommand{\Dbpar}{\ensuremath{D_{\bar{f}}}\xspace}
\newcommand{\Spar}{\ensuremath{S_f}\xspace}
\newcommand{\Sbpar}{\ensuremath{S_{\bar{f}}}\xspace}
\def\normlamsq{|\lambda_{f}|^{2}}
\def\normlambarsq{|\bar{\lambda}_{\bar{f}}|^{2}}
\def\beq{\begin{equation}}
\def\eeq#1{\label{#1}\end{equation}}
\def\eeqn{\end{equation}}

\def\beqa{\begin{eqnarray}}
\def\eeqa#1{\label{#1}\end{eqnarray}}
\def\eeqan{\end{eqnarray}}

\let\bar=\overbar

\def\D{{\cal D}}

\def\Dslash{\not{\hbox{\kern-4pt $D$}}}
\def\dslash{\not{\hbox{\kern-2pt $\del$}}}

\def\msb{{\bar{\ssstyle M \kern -1pt S}}}

\def\ux85 {\mbox{UX85}\xspace}

\ifthenelse{\boolean{uprightparticles}}%
{

 \def\Pmu         {\ensuremath{\upmu}\xspace}

 \def\Ppi         {\ensuremath{\uppi}\xspace}

 \def\Ppsi        {\ensuremath{\uppsi}\xspace}

 \def\PDelta      {\ensuremath{\Delta}\xspace}                 
 \def\PXi      {\ensuremath{\Xi}\xspace}                 
 \def\PLambda      {\ensuremath{\Lambda}\xspace}                 
 \def\PSigma      {\ensuremath{\Sigma}\xspace}                 
 \def\POmega      {\ensuremath{\Omega}\xspace}                 
 \def\PUpsilon      {\ensuremath{\Upsilon}\xspace}                 
 
 \def\PB      {\ensuremath{\mathrm{B}}\xspace}                 
                  
 \def\PD      {\ensuremath{\mathrm{D}}\xspace}

 \def\PJ      {\ensuremath{\mathrm{J}}\xspace}                 
 \def\PK      {\ensuremath{\mathrm{K}}\xspace}

 \def\Pi      {\ensuremath{\mathrm{i}}\xspace}

 \def\Ps      {\ensuremath{\mathrm{s}}\xspace}

}
{

 \def\Pmu         {\ensuremath{\mu}\xspace}

 \def\Ppi         {\ensuremath{\pi}\xspace}

 \def\Ppsi        {\ensuremath{\psi}\xspace}                 
                  
 \mathchardef\PDelta="7101
 \mathchardef\PXi="7104
 \mathchardef\PLambda="7103
 \mathchardef\PSigma="7106
 \mathchardef\POmega="710A
 \mathchardef\PUpsilon="7107
                  
 \def\PB      {\ensuremath{B}\xspace}                 
                  
 \def\PD      {\ensuremath{D}\xspace}

 \def\PJ      {\ensuremath{J}\xspace}                 
 \def\PK      {\ensuremath{K}\xspace}

 \def\Pi      {\ensuremath{i}\xspace}

 \def\Ps      {\ensuremath{s}\xspace}

}

\def\mup        {\ensuremath{\Pmu^+}\xspace}

\def\squark    {\ensuremath{\Ps}\xspace}

\def\pion  {\ensuremath{\Ppi}\xspace}
\def\piz   {\ensuremath{\pion^0}\xspace}

\def\pip   {\ensuremath{\pion^+}\xspace}
\def\pim   {\ensuremath{\pion^-}\xspace}

\def\kaon  {\ensuremath{\PK}\xspace}
  \def\Kbar  {\kern 0.2em\overline{\kern -0.2em \PK}{}\xspace}

\def\Kz    {\ensuremath{\kaon^0}\xspace}
\def\Kzb   {\ensuremath{\Kbar^0}\xspace}
\def\KzKzb {\ensuremath{\Kz \kern -0.16em \Kzb}\xspace}
\def\Kp    {\ensuremath{\kaon^+}\xspace}
\def\Km    {\ensuremath{\kaon^-}\xspace}

\def\KpKm  {\ensuremath{\Kp \kern -0.16em \Km}\xspace}

  \def\Dbar    {\kern 0.2em\overline{\kern -0.2em \PD}{}\xspace}
\def\D       {\ensuremath{\PD}\xspace}

\def\Dz      {\ensuremath{\D^0}\xspace}
\def\Dzb     {\ensuremath{\Dbar^0}\xspace}
\def\DzDzb   {\ensuremath{\Dz {\kern -0.16em \Dzb}}\xspace}
\def\Dp      {\ensuremath{\D^+}\xspace}
\def\Dm      {\ensuremath{\D^-}\xspace}

\def\DpDm    {\ensuremath{\Dp {\kern -0.16em \Dm}}\xspace}

\def\Dstarm  {\ensuremath{\D^{*-}}\xspace}

\def\Dsp     {\ensuremath{\D^+_\squark}\xspace}
\def\Dsm     {\ensuremath{\D^-_\squark}\xspace}

\def\Dss     {\ensuremath{\D^{*+}_\squark}\xspace}

\def\B       {\ensuremath{\PB}\xspace}
  \def\Bbar    {\kern 0.18em\overline{\kern -0.18em \PB}{}\xspace}
\def\Bb      {\ensuremath{\Bbar}\xspace}
 
\def\Bz      {\ensuremath{\B^0}\xspace}
\def\Bzb     {\ensuremath{\Bbar^0}\xspace}
\def\Bu      {\ensuremath{\B^+}\xspace}
\def\Bub     {\ensuremath{\B^-}\xspace}
\def\Bp      {\ensuremath{\Bu}\xspace}
\def\Bm      {\ensuremath{\Bub}\xspace}

\def\Bs      {\ensuremath{\B^0_\squark}\xspace}
\def\Bsb     {\ensuremath{\Bbar^0_\squark}\xspace}

\def\jpsi     {\ensuremath{{\PJ\mskip -3mu/\mskip -2mu\Ppsi\mskip 2mu}}\xspace}

  \def\Y#1S{\ensuremath{\PUpsilon{(#1S)}}\xspace}

\def\Lbar {\ensuremath{\kern 0.1em\overline{\kern -0.1em\PLambda}}\xspace}

\def\to                 {\ensuremath{\rightarrow}\xspace}

\def\CP                {\ensuremath{C\!P}\xspace}

\def\AT#1     {\ensuremath{A_{\mathrm{T}}^{#1}}\xspace}           

\def\C#1      {\ensuremath{\mathcal{C}_{#1}}\xspace}                       
\def\Cp#1     {\ensuremath{\mathcal{C}_{#1}^{'}}\xspace}                    
\def\Ceff#1   {\ensuremath{\mathcal{C}_{#1}^{\mathrm{(eff)}}}\xspace}        
\def\Cpeff#1  {\ensuremath{\mathcal{C}_{#1}^{'\mathrm{(eff)}}}\xspace}       
\def\Ope#1    {\ensuremath{\mathcal{O}_{#1}}\xspace}                       
\def\Opep#1   {\ensuremath{\mathcal{O}_{#1}^{'}}\xspace}                    

\newcommand{\tev}{\ensuremath{\mathrm{\,Te\kern -0.1em V}}\xspace}
\newcommand{\gev}{\ensuremath{\mathrm{\,Ge\kern -0.1em V}}\xspace}
\newcommand{\mev}{\ensuremath{\mathrm{\,Me\kern -0.1em V}}\xspace}
\newcommand{\kev}{\ensuremath{\mathrm{\,ke\kern -0.1em V}}\xspace}
\newcommand{\ev}{\ensuremath{\mathrm{\,e\kern -0.1em V}}\xspace}
\newcommand{\gevc}{\ensuremath{{\mathrm{\,Ge\kern -0.1em V\!/}c}}\xspace}
\newcommand{\mevc}{\ensuremath{{\mathrm{\,Me\kern -0.1em V\!/}c}}\xspace}
\newcommand{\gevcc}{\ensuremath{{\mathrm{\,Ge\kern -0.1em V\!/}c^2}}\xspace}
\newcommand{\gevgevcccc}{\ensuremath{{\mathrm{\,Ge\kern -0.1em V^2\!/}c^4}}\xspace}
\newcommand{\mevcc}{\ensuremath{{\mathrm{\,Me\kern -0.1em V\!/}c^2}}\xspace}

\def\gsim{{~\raise.15em\hbox{$>$}\kern-.85em
          \lower.35em\hbox{$\sim$}~}\xspace}
\def\lsim{{~\raise.15em\hbox{$<$}\kern-.85em
          \lower.35em\hbox{$\sim$}~}\xspace}

\newcommand{\Real}{\ensuremath{\mathcal{R}e}\xspace}
\newcommand{\Imag}{\ensuremath{\mathcal{I}m}\xspace}

\def\sWeights{\mbox{\em sWeights}}

\def\tell1  {TELL1\xspace}
\def\ukl1   {UKL1\xspace}

\def\dzdzb{{~\raise.85em\hbox{{\tiny{(}\textemdash\tiny{)}}}\kern-1.05em
          \lower0.0em\hbox{$D^0$}~}\xspace}
\def\bsbsb{{~\raise.85em\hbox{{\tiny{(}\textemdash\tiny{)}}}\kern-1.05em
          \lower0.0em\hbox{$B_s^0$}~}\xspace}

\def\br{{\cal{B}}}

\def\bstodspipipi{\Bsb\to\Dsp\pi^-\pi^+\pi^-}
\def\bstodskpipi{\Bsb\to\Dsp K^-\pi^+\pi^-}

\def\btodskpipi{\Bzb\to\Dsp K^-\pi^+\pi^-}

\def\bstodsstarpipipi{\Bsb\to\Dss\pi^-\pi^+\pi^-}

\def\bstodspi{\Bsb\to D_s^+\pi^-}
\def\bstodsk{\Bsb\to D_s^{+}K^{-}}

\def\btodzerok{\Bm\to D\Km}

\def\Title#1{\begin{center} {\Large {\bf #1} } \end{center}}

\begin{document}

\Title{Measurement of the CP observables in $\bstodsk$ and first observation of
  $\Bzb_{(s)}\to\Dsp\Km\pip\pim$ and $\Bsb\to D_{s1}(2536)^+\pim$}

\bigskip\bigskip


\begin{raggedright}  

{\it Steven R. Blusk\index{Blusk, S.}\\
Department of Physics\\
Syracuse University\\
Syracuse, NY 13244, USA}\\
Proceedings of CKM 2012, the 7$^{th}$ International Workshop on the CKM
Unitarity Triangle, University of Cincinnati, USA, 28th September - 2 October 2012
\bigskip\bigskip
\end{raggedright}

\linenumbers

\section{Introduction}

A central goal of flavor physics is to measure the angle 
${\gamma\equiv {\rm arg}\left(-{V_{\rm ub}^*V_{\rm ud}\over V_{\rm cb}^*V_{\rm cd}}\right)}$
in the Cabibbo-Kobayashi-Maskawa (CKM)~\cite{Cabibbo:1963yz,Kobayashi:1973fv} mixing matrix, which is
currently known to a precision of about 10-12$^{\rm o}$~\cite{combfits}. The theoretically cleanest
methods employ $B\to DK$ decays, where the sensitivity to $\gamma$ results from the interference
between $b\to c$ and $b\to u$ transitions. Since both transitions are $\mathcal{O}(\lambda^3$) in the Wolfenstein
parameter~\cite{Wolfenstein:1983yz}, large CP violating asymmetries are expected.
One powerful class of methods utilize $\btodzerok$
where the $D$ is detected in either a CP eigenstate~\cite{glw}, a flavor-specific mode~\cite{ads}, or a multi-body decay~\cite{ggsz}. 
An advantage of these decays is that they do not require knowledge of the $b$-hadron flavor at 
production (flavor tagging), and only rely on measuring the time integrated rates. Another powerful method to extract $\gamma$ is to perform a
time-dependent analysis of $\bstodsk$~\cite{aleksan,fleischer,fleischer2} and $\bstodskpipi$.
Time-dependent analyses
of $\bstodsk$$(\pip\pim)$ are only possible at hadron colliders, and are a unique capability of LHCb.

The time-dependent decay rates of $\Bs$ and $\Bsb$ to a flavor-specific final state, $f=\Dsp\Km$, is given by:
\begin{eqnarray}
\label{eq:decay_rates_1}
\frac{d\Gamma_{B^0_s\rightarrow f}(t)}{dt} & = \frac{1}{2}|A_f|^2 (1+|\lambda_f|^2) e^{-\Gamma_s t} &
\left[\cosh\left(\frac{\Delta\Gamma_s t}{2}\right) + D_f \sinh\left(\frac{\Delta\Gamma_s t}{2}\right)\right. \nonumber \\
& & \left.\phantom{\Big(}
+C_f \cos\left(\Delta m_s t\right) - S_f \sin\left(\Delta m_s t\right) \right. \bigg{]},\\
 & & \nonumber \\
\frac{d\Gamma_{\bar{B}^0_s\rightarrow f}(t)}{dt} & = \frac{1}{2}|A_f|^2 \left|\frac{p}{q}\right|^2 (1+|\lambda_f|^2) e^{-\Gamma_s t} &
\left[\cosh\left(\frac{\Delta\Gamma_s t}{2}\right) + D_f \sinh\left(\frac{\Delta\Gamma_s t}{2}\right)\right. \nonumber \\
& & \left.\phantom{\Big)}
- C_f \cos\left(\Delta m_s t\right) + S_f \sin\left(\Delta m_s t\right) \right. \bigg{]},\\
& & \nonumber
\label{eq:decay_rates_2}
\end{eqnarray}
where $A_f$ is the decay amplitude $A(B^{0}_{s}\to f)$ and
$\lambda_{f} = (q/p)(\bar{A}_{f}/A_{f})=|\lambda_f|e^{i(\Delta-(\gamma-2\beta_s))}$. Here, $|\lambda_f|$
and $\Delta$ are the relative magnitude and strong phase difference between the $b\to u$ and $b\to c$ transitions,
and $2\beta_s$ is the phase of $\Bs$ mixing.
The complex coefficients $p$ and $q$ relate the $B^0_s$ meson mass eigenstates, $B_{H,L}$, to the 
flavor eigenstates, $B^0_s$ and $\bar{B}^0_s$ via:
\begin{equation}
\begin{array}{c}
B_L = pB^0_s+q\bar{B}^0_s \\
B_H = pB^0_s-q\bar{B}^0_s
\end{array}
,\qquad |p|^2+|q|^2=1 \,.
\label{eq:mixing}
\end{equation}
Similar equations can be written for the \CP-conjugate decays, replacing
$A_f$ by $\bar{A}_{\bar{f}} = A(\Bsb\to\bar{f})$, 
$\lambda_f$ by $\bar{\lambda}_{\bar{f}} = (p/q)(A_{\bar{f}}/\bar{A}_{\bar{f}})$,
$|p/q|^2$ by $|q/p|^2$, $C_f$ by \Cbpar, \Spar by \Sbpar, and \Dpar by \Dbpar.
The \CP asymmetry observables $C_f$, $S_f$, $D_f$, $C_{\bar{f}}$,
$S_{\bar{f}}$ and $D_{\bar{f}}$ are then given by
\begin{eqnarray}
C_f = C_{\bar{f}} = \frac{ 1 - \normlamsq }{ 1 + \normlamsq } \,,\quad  \quad
S_f = \frac{ 2 \Imag(\lambda_f) }{ 1 + \normlamsq } \,,\quad  \quad
D_f = \frac{ 2 \Real(\lambda_f) }{ 1 + \normlamsq } ,\quad \nonumber \\
S_{\bar{f}} = \frac{ 2 \Imag(\bar{\lambda}_{\bar{f}}) }{ 1 + \normlambarsq } \,,\quad  \quad
D_{\bar{f}} = \frac{ 2 \Real(\bar{\lambda}_{\bar{f}}) }{ 1 + \normlambarsq } \,.
\label{eq:asymm_obs}
\end{eqnarray}
Since CP violation in mixing is expected to be below the percent level, it follows that 
$|q/p| = 1$,  $|\lambda_f| = |\bar{\lambda}_{\bar{f}}|$, and consequently $C_f = C_{\bar{f}}$.
Thus there are five observables that depend on the 3 physics parameters of interest: $|\lambda_f|$, 
$\Delta$ and $\gamma-2\beta_s$.
Similar expressions are applicable to $\bstodskpipi$, however, there is a potential dilution due to the varying 
strong phase across the $\Dsp\Km\pip\pim$ Dalitz plane.

In this article, we present the first measurements of these five CP observables. First observations of 
the $\bstodskpipi$, $\btodskpipi$ and $\Bb_s\to D_{s1}(2536)^+\pim$ decays are also presented, along with measurements of
their relative branching fractions. All results are based on 1.0~fb$^{-1}$ of integrated luminosity recorded in 2011 by 
the LHCb experiment. More detailed documentation of the $\bstodsk$ and $\Bzb_{(s)}\to\Dsp\Km\pip\pim$
analyses can be found in Refs.~\cite{bstodskConf} and~\cite{bstodskpipiPaper}, respectively.

\section{Event Selection}

\begin{figure}[htb]
\begin{center}
\epsfig{file=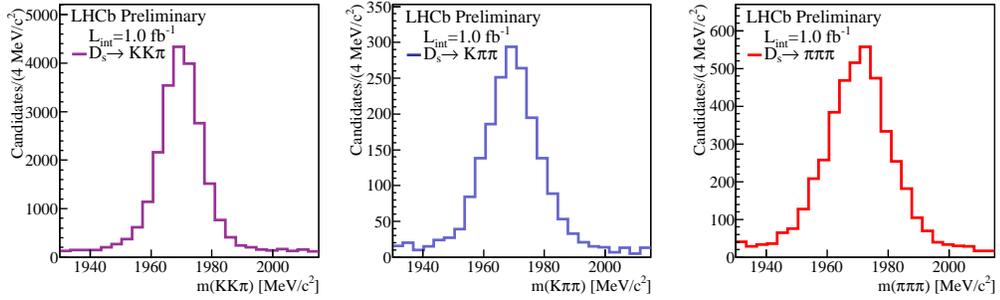,width=0.9\linewidth}
\caption{Invariant mass distributions for $\Dsp$ candidates in the $\bstodspi$ data sample, for (left) $\Kp\Km\pip$, (middle) $\Kp\pim\pip$, and 
(right) $\pip\pim\pip$ final states.}
\label{fig:DsMassPlots}
\end{center}
\end{figure}

Signal $\Dsp$ candidates are formed by reconstructing $\Dsp\to\Kp\Km\pip$, $\Dsp\to\pip\pim\pip$ and $\Dsp\to\Kp\pim\pip$.
For the $\Bzb_{(s)}\to\Dsp\Km\pip\pim$ and $\bstodspipipi$ candidates, only the $\Dsp\to\Kp\Km\pip$ decay is considered.
The $\Dsp$ candidates are required to form a good quality vertex, be spacially well separated from any primary vertex (PV), and
have an invariant mass consistent with the known $\Dsp$ mass (within about 3 times the mass resolution). 
Multivariate selection algorithms
are employed to suppress the combinatorial background, and typically have a signal efficiency of 80-90\% while
rejecting about 85\% of the combinatorial background. Invariant mass distributions for $\Dsp$ candidates are
shown in Fig.~\ref{fig:DsMassPlots} for the higher signal yield $\bstodspi$ decay, showing that clean signals
are achievable even in the suppressed $\Dsp$ decay modes.
Tighter particle identification requirements are applied to the $\Km$ or $\Km\pip\pim$ recoiling from the $\Dsp$ to
suppress cross-feed from the favored $\bstodspi$ and $\bstodspipipi$ decays. For the $\bstodspipipi$ and $\bstodskpipi$ decays,
the invariant mass of the $\pim\pip\pim$ and $\Km\pip\pim$ systems are restricted to be below 3000~\mevcc.

\section{Analysis of $\bstodspi$ and $\bstodsk$}

\begin{figure}[htb]
\begin{center}
\epsfig{file=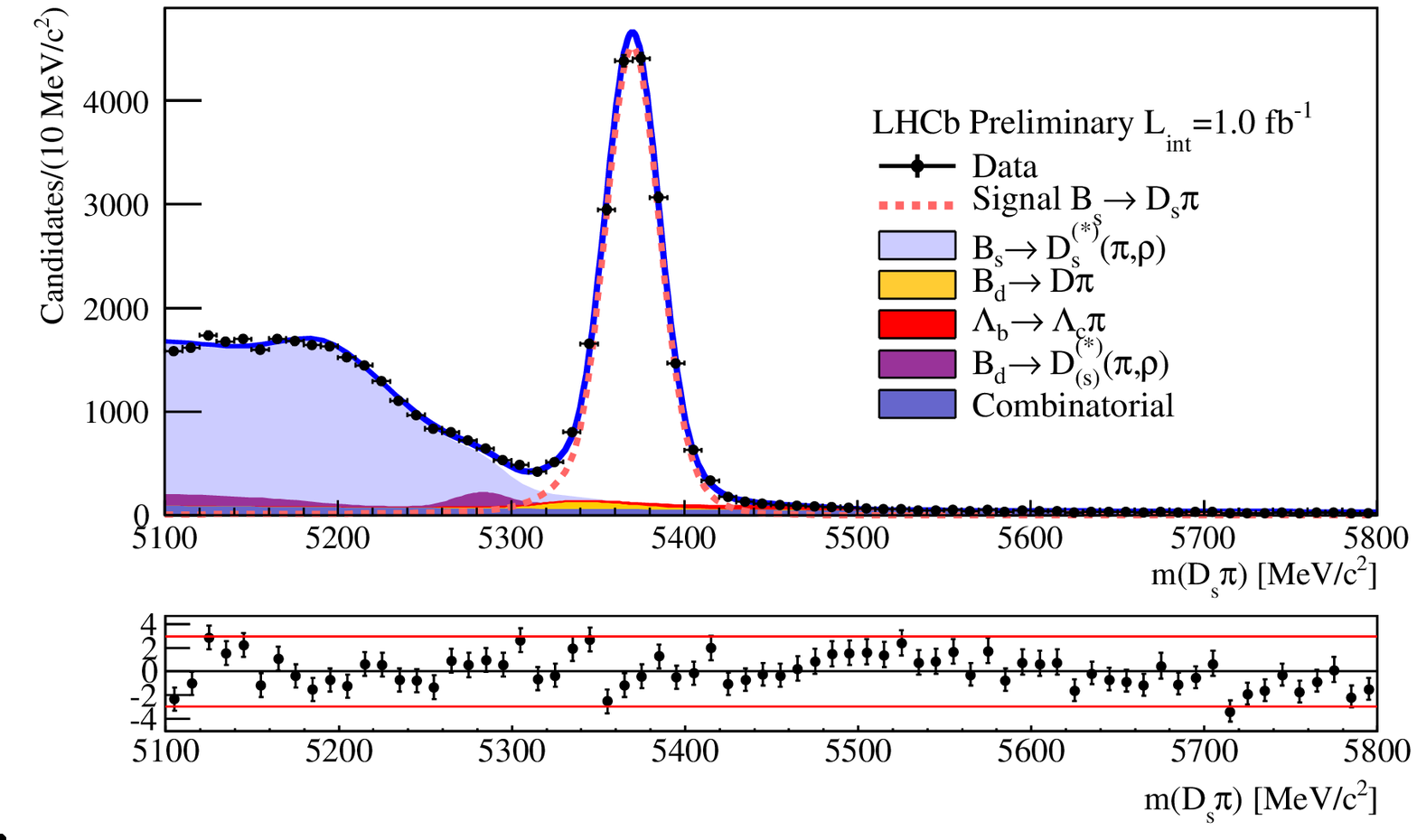,width=0.9\linewidth,clip=,bb=20 20 567 319}
\caption{Invariant mass distributions $\bstodspi$ candidates. The signal component is indicated by the dashed curve,
and the backgrounds are indicated by the various color-filled (shaded, in B/W) curves.}
\label{fig:Bs2DsPiMassPlot}
\end{center}
\end{figure}

\begin{figure}[htb]
\begin{center}
\epsfig{file=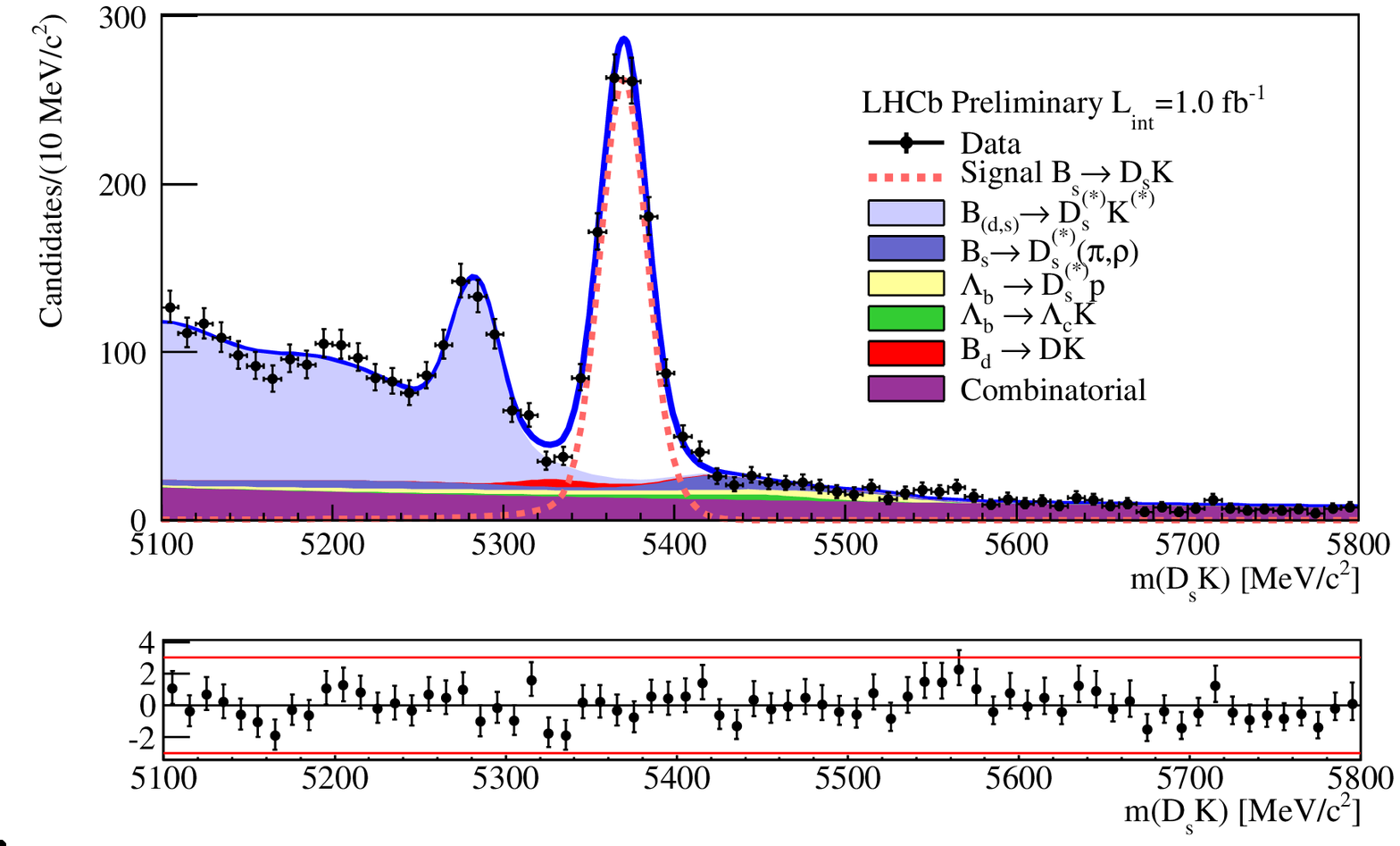,width=0.9\linewidth,clip=,bb=20 20 567 319}
\caption{Invariant mass distributions $\bstodsk$ candidates. The signal component is indicated by the dashed curve,
and the backgrounds are indicated by the various color-filled (shaded, in B/W) curves.}
\label{fig:Bs2DsKMassPlot}
\end{center}
\end{figure}

The invariant mass distributions for $\bstodspi$ and $\bstodsk$ are shown in Figs.~\ref{fig:Bs2DsPiMassPlot} 
and~\ref{fig:Bs2DsKMassPlot}. All three $\Dsp$ decay modes have approximately equal $\Bs$ mass resolutions, and
are summed together in these distributions. The signal shape is modeled as
the sum of two Crystal Ball~\cite{Skwarnicki:1986xj} functions, with one exponential tail on each side of the $\Bsb$ signal peak.
A number of specific backgrounds, due to either a missed particle (e.g. $\Bsb\to\Dsp\rho^-$, with the $\piz$ undetected), 
a misidentified particle (e.g. $\bstodspi$ reconstructed as $\bstodsk$), or both (e.g. $\Bsb\to\Dsp\rho^-$ reconstructed as
$\bstodsk$) are accounted for using either data or simulation to model the shape of these backgrounds. 
From an unbinned extended maximum likelihood fit, $27,965\pm395$ $\bstodspi$ and $1390\pm98$ $\bstodsk$
signal events are selected.
The CP parameters are obtained by a fit to the decay time distribution of the $\bstodsk$ signal candidates.
Two methods have been developed. The first, referred to as {\it sFit}, uses \sWeights~\cite{Pivk:2004ty} obtained from 
the $\bstodsk$ mass fit to statistically subtract the background contribution. The second method, referred to as {\it cFit}, is
a conventional two-dimensional fit to the reconstructed mass and decay time. The advantage of the first method is that
there is no need to model the time distribution of all the backgrounds, as they are statistically removed via the
\sWeights. The statistical subtraction, as presented here, uses events in the full mass fit region, and the subtraction of this
background leads to a larger statistical uncertainty than if just a narrow signal region is used. 
For this reason, the second method is expected to give a smaller 
statistical uncertainty; however it requires an accurate model of the time distributions of the backgrounds
that enter into the signal region.
For the analysis presented here, the {\it sFit} provides the nominal result, and the {\it cFit} is used as a cross-check.

The measurement of the CP parameters in $\bstodsk$ requires a fit to the time-dependent decay rates. The fit accounts for
(i) the acceptance versus reconstructed decay time, (ii) the decay time resolution, and (iii) the effective tagging efficiency.
The functional form of the acceptance function is determined from simulated $\bstodspi$, and its parameters are
determined in a fit to $\bstodspi$ data, where the $\Bs$ lifetime and mixing frequency, $\Delta m_s$, are
fixed to 1.51 ps and 17.69~ps$^{-1}$~\cite{hfag}, respectively. The average decay time resolution is about 50~fs, and
is modeled by the sum of three
Gaussian functions, whose parameters are determined from simulation. The Gaussian width parameters obtained
from simulation are scaled up by 1.15 to account for better resolution in the simulation than in data; this factor  is 
determined by comparing the width of the zero decay time component of prompt $\Dsp$ plus one random track in 
data and simulation. For the flavor tagging, only {\it opposite side} (OS) taggers are currently used. These
algorithms exploit the correlation in flavor between the signal $b$ hadron at production, and the other $b$ hadron
in the event (referred to as the tag-$b$). In particular, the charge of either an electron, a muon, or a kaon
that does not come from any $pp$ interaction vertex (or the signal $b$), or the charge of another secondary vertex in the event, 
provide information on the flavor of the tag-$b$ hadron. Because $b\bar{b}$ are produced in pairs, this translates into a flavor determination of
the signal $\Bs$. The OS flavor tagging algorithm was initially tuned using simulated decays, and then re-optimized
and calibrated to obtain the largest effective tagging efficiency
using the self-tagging $\Bp\to\jpsi\Kp$ and $\Bz\to\Dstarm\mup\nu$ decays in data.
In general, the performance of the OS tagging algorithms are independent of the
signal $b$-hadron decay, and have a combined effective tagging efficiency of $\epsilon D^2 = 1.90\%$
for $\bstodsk$. Further details of the tagging algorithms can be found in Ref~\cite{tagging}.

\begin{figure}[htb]
\begin{center}
\epsfig{file=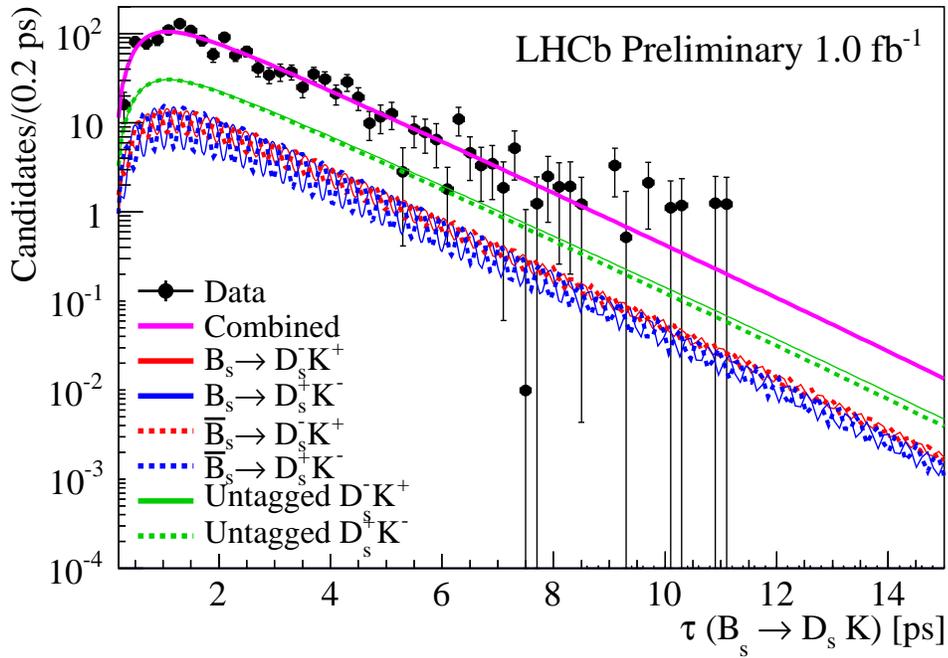,width=0.9\linewidth}
\caption{Distribution of reconstruct decay time for $\bstodsk$ signal decays (points with error bars),
along with the results of the fit. Projections of the decay rates versus the decay time for the four
possible flavor tagged decays, and the two untagged decays.}
\label{fig:Bs2DsKTimePlot}
\end{center}
\end{figure}

In the fit to $\bstodsk$, the following parameters are fixed: 
$\Delta m_s=17.69$~ps$^{-1}$, $\tau_{B_s}=1.51$~ps and 
$\Delta\Gamma_s\equiv\Gamma_{s,L}-\Gamma_{s,H}=0.105$~ps$^{-1}$~\cite{hfag}. 
About 60\% of the $\bstodsk$ candidates have no flavor tag;
the time-dependent decay rates for these untagged decays is given by the sum of the two expressions in  Eq.~\ref{eq:decay_rates_2},
and the sensitivity to $\gamma$ enters through the hyperbolic sine term. The decay time
distribution of $\bstodsk$ signal decays and projections of the fitted 
are shown in Fig.~\ref{fig:Bs2DsKTimePlot}. The projections show the four possible
tagged decays, $\Bs\to D_s^{\pm}K^{\mp}$ and $\Bsb\to D_s^{\pm}K^{\mp}$, as well as the
untagged time-dependent decay rates $(\Bs,\Bsb)\to\Dsm\Kp$ and $(\Bs,\Bsb)\to\Dsp\Km$. 
The fitted values for the CP parameters are
\begin{align*}
\Cpar  &= \phantom{-}1.01 \pm 0.50 \pm 0.23\,, \\
\Spar  &= -1.25 \pm 0.56 \pm 0.24\,, \\
\Sbpar &= \phantom{-}0.08 \pm 0.68 \pm 0.28\,, \\
\Dpar  &= -1.33 \pm 0.60 \pm 0.26\,, \\
\Dbpar &= -0.81 \pm 0.56 \pm 0.26\,,
\end{align*}
where the first uncertainties are statistical and the second are systematic. Several sources of
systematic uncertainty have been considered. The dominant sources are due to the precision on the effective
flavor tagging efficiency (0.16$\sigma_{\rm stat}$-0.23$\sigma_{\rm stat}$), variations in the parameters
that are fixed in the default fits (0.15$\sigma_{\rm stat}$-0.42$\sigma_{\rm stat}$), and the correlation
between the mass of specific backgrounds and their reconstructed decay time (0.08$\sigma_{\rm stat}$-0.27$\sigma_{\rm stat}$),
where these uncertainties are expressed as a fraction of the statistical error.
These are the first measurements of the CP parameters in $\bstodsk$. With additional data and analysis
refinements, reduction in both the statistical and systematic uncertainties are expected.

\section{First observation of $\Bb_{s}\to\Dsp\Km\pip\pim$ and $\Bsb\to D_{s1}(2536)^+\pim$}

The decay $\bstodskpipi$ can be analyzed in a similar way to $\bstodsk$ to measure
the weak phase $\gamma$. While this decay has not yet been observed, if one uses
$\Bzb$ and $\Bm$ decays as a guide, it would naively be expected that its branching fraction
is 1.5-2.0 times larger than $\bstodsk$, making this a potentially attractive decay 
mode to explore. The first step in such an analysis is to firmly establish an observation
of this decay and measure its branching fraction (here, relative to $\bstodspipipi$). While 
searching for this decay, the decay $\Bzb\to\Dsp\Km\pip\pim$ is also observed and its 
branching fraction is measured relative to $\bstodskpipi$. 

\begin{figure}[htb]
\begin{center}
\epsfig{file=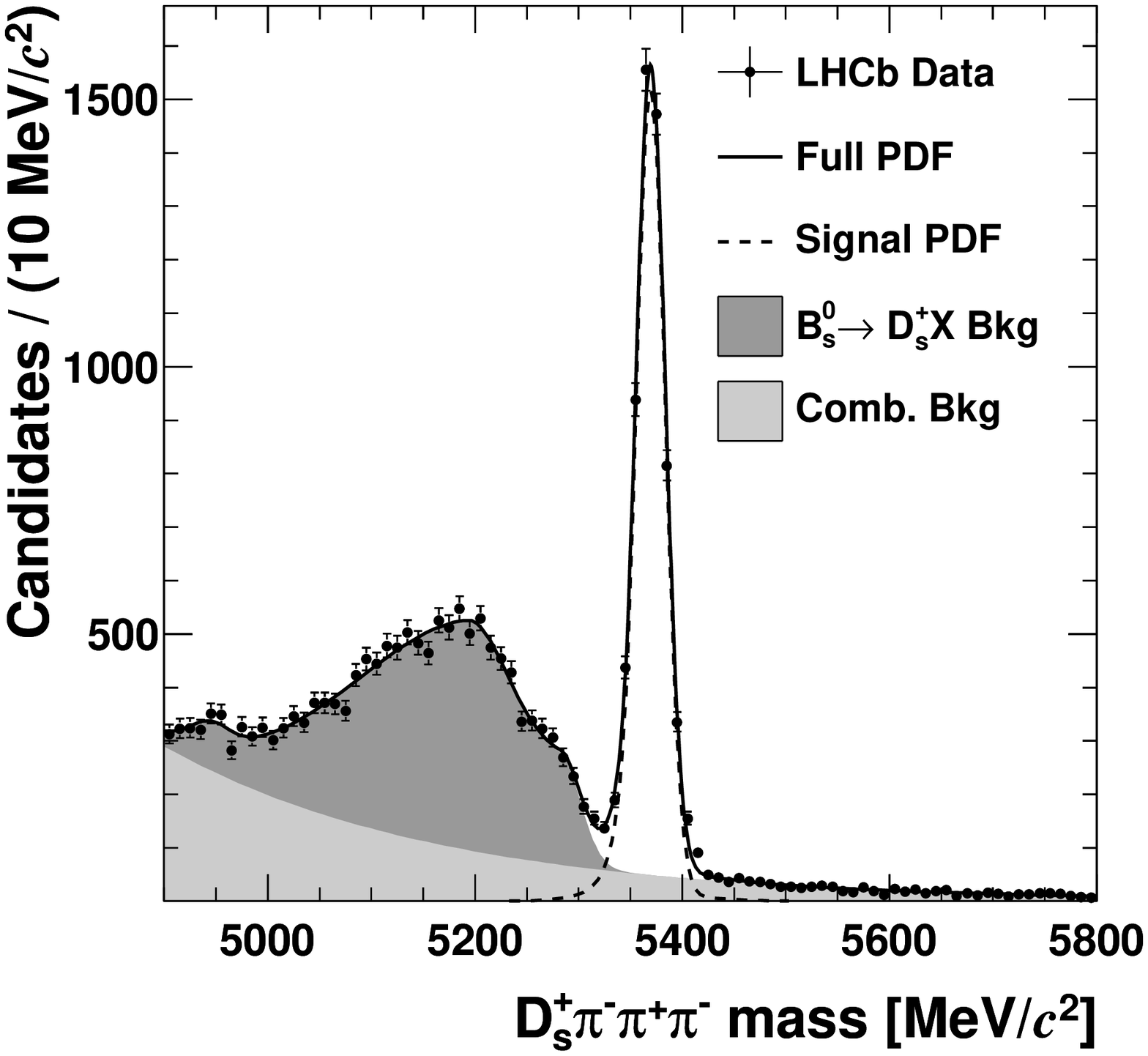,width=0.48\linewidth}
\epsfig{file=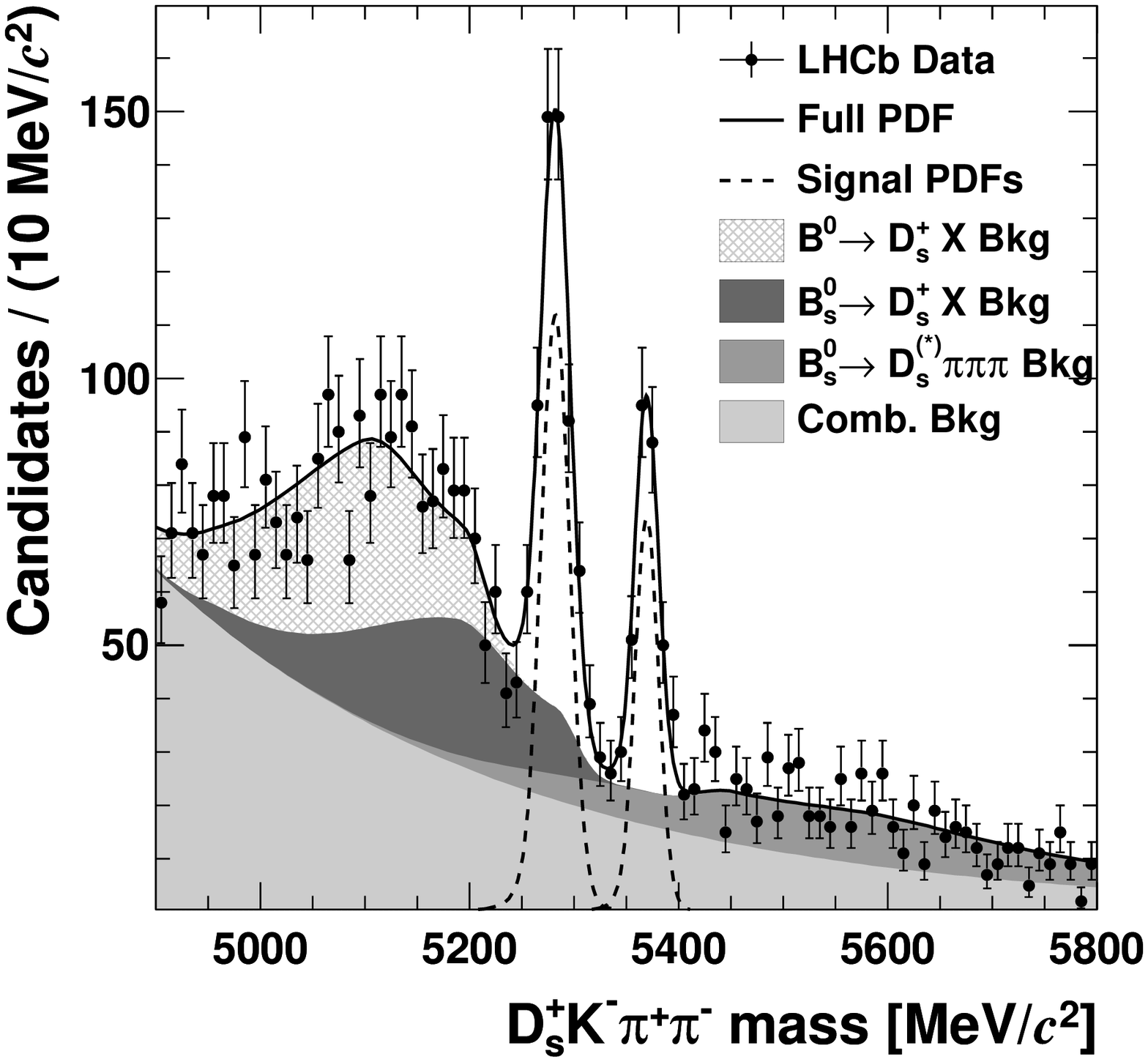,width=0.48\linewidth}
\caption{Invariant mass distribution for (left) $\bstodspipipi$ candidates and 
(right) $\Bzb_{(s)}\to\Dsp\Km\pip\pim$ candidates. The fitted signal (dashed lines) 
and background shapes (shaded/hatched regions) are shown, as described in the text.}
\label{fig:bsDecays}
\end{center}
\end{figure}

With the previously defined selections, Fig.~\ref{fig:bsDecays} shows the
invariant mass distributions for (left) $\bstodspipipi$ candidates and 
(right) $\Bzb_{(s)}\to\Dsp\Km\pip\pim$ candidates. Significant $\Bsb$ signals are
seen in both spectra, and a $\Bzb$ signal is seen in the $\Dsp\Km\pip\pim$ mass 
distribution. The main sources of background are $\bstodsstarpipipi$ (to $\bstodspipipi$),
and $\bstodspipipi$,  $\bstodsstarpipipi$, and $\Bzb_{(s)}\to\Dss\Km\pip\pim$ 
(to $\Bzb_{(s)}\to\Dsp\Km\pip\pim$). Their shapes are taken from simulation, 
with parameters that are allowed to vary within their uncertainties. 
Yields of $5683\pm83$ $\bstodspipipi$, $216\pm21$ $\bstodskpipi$ and 
$402\pm33$ $\btodskpipi$ are observed. After correcting for the relative efficiencies, 
the ratio of branching fractions are measured to be
\begin{align*}
{\br(\bstodskpipi)\over\br(\bstodspipipi)} &= (5.2\pm0.5\pm0.3)\times10^{-2} \\
{\br(\btodskpipi)\over\br(\bstodskpipi)} &= 0.54\pm 0.07\pm0.07, \\
\end{align*}
\noindent \noindent where the uncertainties are statistical and systematic, respectively.
These are the first observations of these decays. Since $\bstodspipipi$ has a branching fraction
that is about twice as large as $\bstodspi$, and $\br(\bstodsk)\sim0.09\times\br(\bstodspi)$~\cite{pdg}, it
follows that $\br(\bstodspipipi)$ is at least as large as $\br(\bstodspi)$, or as much as 
50\% larger. The $\br(\Bzb\to\Dsp\Km\pip\pim)$ is also sizeable, and is likely dominated by
contributions where an extra $s\bar{s}$ pair is produced in addition to the weak decay 
(see Ref.~\cite{bstodskpipiPaper} for more details).

\begin{figure}[htb!]
\begin{center}
\epsfig{file=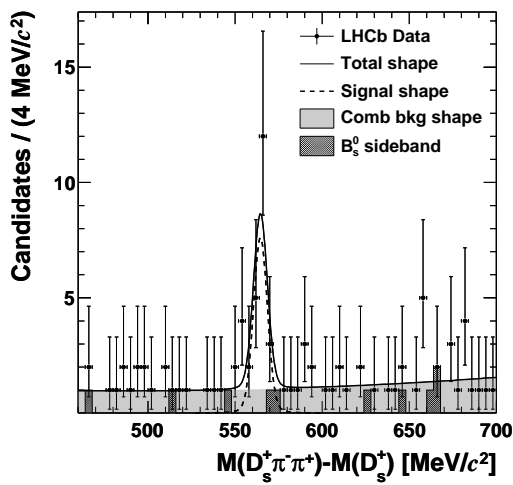,width=0.48\linewidth}
\end{center}
\caption{Distribution of the difference in invariant mass, $M(\Dsp\pim\pip)-M(\Dsp)$, using
$\bstodspipipi$ candidates within 40~\mevcc of the known $\Bs$ mass (points) and in the upper $\Bs$
mass sidebands (filled histogram). The fit to the distribution is shown, as described in the text.}
\label{fig:Bs2Ds1Pi}
\end{figure}

The $\bstodspipipi$ decay has also been analyzed to search for intermediate excited $D_{sj}$
states. For $\bstodspipipi$ candidates within 40~\mevcc of the $\Bsb$ signal peak, 
the mass difference, 
$\Delta M \equiv M(\Dsp\pim\pip)-M(\Dsp)$ is computed for both $\pim\pip$ mass combinations. The resulting
mass difference spectrum is shown in Fig.~\ref{fig:Bs2Ds1Pi}. The signal is fit with a
Breit-Wigner convolved with a Gaussian resolution function whose width is fixed to the expected
$\Delta M$ resolution. A signal of $20.0\pm5.1$ events is observed with a $\Delta M$ value and width 
consistent with the $D_{s1}(2536)^+$ state. Applying corrections for the relative efficiency, the 
ratio of branching fractions is measured to be
\begin{align*}
{\br(\Bsb\to D_{s1}(2536)^+\pim,~D_{s1}^+\to\Dsp\pim\pip)\over\br(\bstodspipipi)} = (4.0\pm1.0\pm0.4)\times10^{-3}. \\
\end{align*}

\noindent The excess of events is 5.9 standard deviations over the expected background, thus establishing
the first observation of this decay. 

\section{Summary}
First measurements of the CP observables in the $\bstodsk$ decay have been reported. With
the larger data sample recorded in 2012, and the larger data set anticipated in the future,
this decay will contribute significantly to the determination of the weak phase $\gamma$. First observations
of the $\bstodskpipi$ and $\btodskpipi$ are also reported. The former can be used in a similar
way to $\bstodsk$ to extract $\gamma$. After including $\Dsp\to\pip\pim\pip$ and $\Dsp\to\Km\pip\pim$ decays,
and reoptimizing the selection for $\bstodskpipi$ only, the yield in this mode more than doubles with
a comparable signal-to-background. The yield in this mode is therefore expected to have about 35-40\% of
that obtained in $\bstodsk$. The $\Bsb\to D_{s1}(2536)^+\pim$ decay is also observed for the first time, and
its branching fraction relative to $\bstodspipipi$ is presented.

\section*{Acknowledgements}
I gratefully acknowledge support from the National Science Foundation, which makes this research possible.

\end{document}